\pdfoutput=1
\RequirePackage{ifpdf}

\documentclass[cits,11pt,a4paper]{article}
\usepackage{jheppub}

\usepackage{graphicx}
\usepackage{amsmath}
\usepackage{lineno}

\usepackage{stmaryrd}
\usepackage[page]{appendix}

\def\bfseries{\fontseries \bfdefault \selectfont \boldmath}

\title{Comment on ``Damping of neutrino oscillations, decoherence and the lengths of neutrino wave packets''}

\author[1,a]{B.J.P.~Jones,\note[a]{Corresponding author.}}
\affiliation[1]{
Department of Physics, University of Texas at Arlington, Arlington, TX 76019, USA}
\abstract{We point out three apparent inconsistencies in the treatment of oscillation
coherence from reactor neutrino and source neutrino experiments in
recent paper \cite{akhsmirn}. First, that the dependence of the oscillation
probability upon the subsequent interactions of entangled recoil particles
implies causality violations and in some situations superluminal signaling;
second, that integrating over a non-orthogonal basis for the entangled
recoil leads to unphysical effects; and third, that the question of
what interactions serve to measure the position of the initial state
particle remains ambiguous. These points taken together appear to
undermine the claim made therein that the effects of wave packet separation
must be strictly unobservable in reactor and radioactive source based
neutrino experiments.
}

\begin{document}
\maketitle
\section{Causality violations\label{sec:Causality-violations}}

Ref.~\cite{akhsmirn} considers the wave packet length resultant coherence
properties of oscillating neutrinos emerging from nuclear decays.
The formalism used there distinguishes between two scenarios: the
decay product(s) accompanying the neutrino escape undetected, or they
interact with source material, representing a measurement. The frequency
of interaction of the decay product(s), either electron or nuclear
recoil in this case, with the material is considered as a timescale
over which neutrino emission coherence is interrupted. It is thus
proposed to be an influential timescale (and hence distance scale)
upon the neutrino wave packet length, and in some cases the critical
timescale that determines oscillation coherence. 

\begin{figure}
\begin{centering}
\includegraphics[width=0.6\columnwidth]{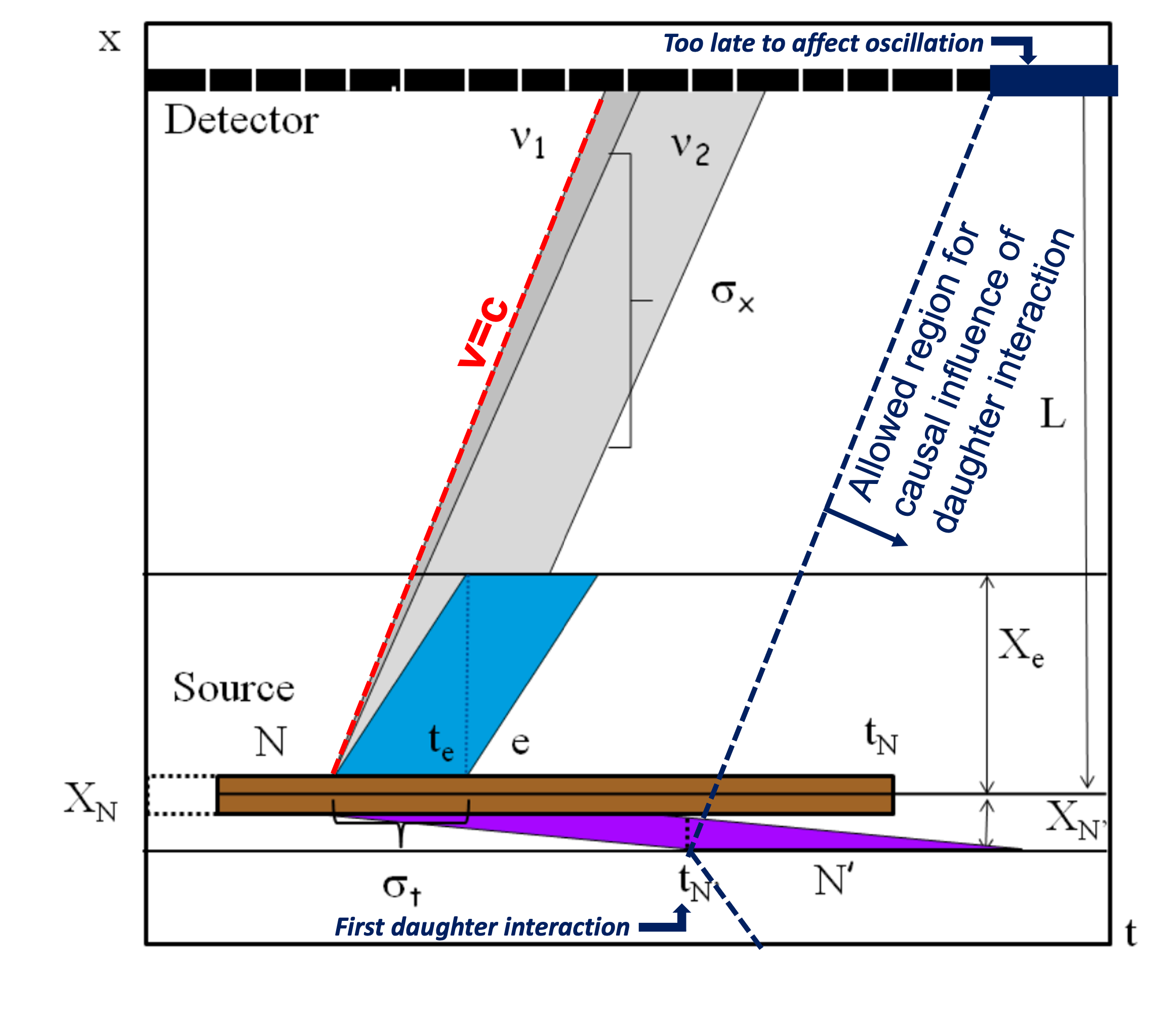}
\par\end{centering}
\caption{Light cone of the nuclear recoil interaction overlaid on the spacetime
diagram from Ref.~\cite{akhsmirn}. If interactions of the recoil
at time $t_{N'}$ may influence oscillations of the neutrinos at $L$,
causality is violated. \label{fig:Light-cone-of}}
 
\end{figure}

While superficially plausible, this picture presents a serious problem,
in that it implies causality violations that in principle allow for
superluminal propagation of signals between recoil and neutrino. An
event happening to nuclear or electronic recoil at time $t_{R}$ has
a corresponding future light-cone (the region of spacetime in which
it can causally influence any observable at the detector) which meets
the detector location $L$ at time $t_{D}=t_{R}+\frac{L}{c}$. Given
that the recoil moves slowly relative to the neutrino,
the light cone emanating from its interaction point almost certainly reaches the detector baseline when
the neutrino detection process is already complete. This is shown
overlaid on the spacetime diagram of Ref.~\cite{akhsmirn} in Fig.
\ref{fig:Light-cone-of}, using the nuclear recoil for illustration.
If the neutrino oscillation probability is indeed influenced by detection-or-not
or interaction-or-not of the recoiling particles, causality has been
violated. An equivalent statement is that since the entangled partners
emerge from a common vertex with the neutrino and travel at less than
$c$, they are space-like separated from it and so their subsequent
behavior cannot causally impact the oscillation phenomenology. 

\begin{figure}
\begin{centering}
\includegraphics[width=0.9\columnwidth]{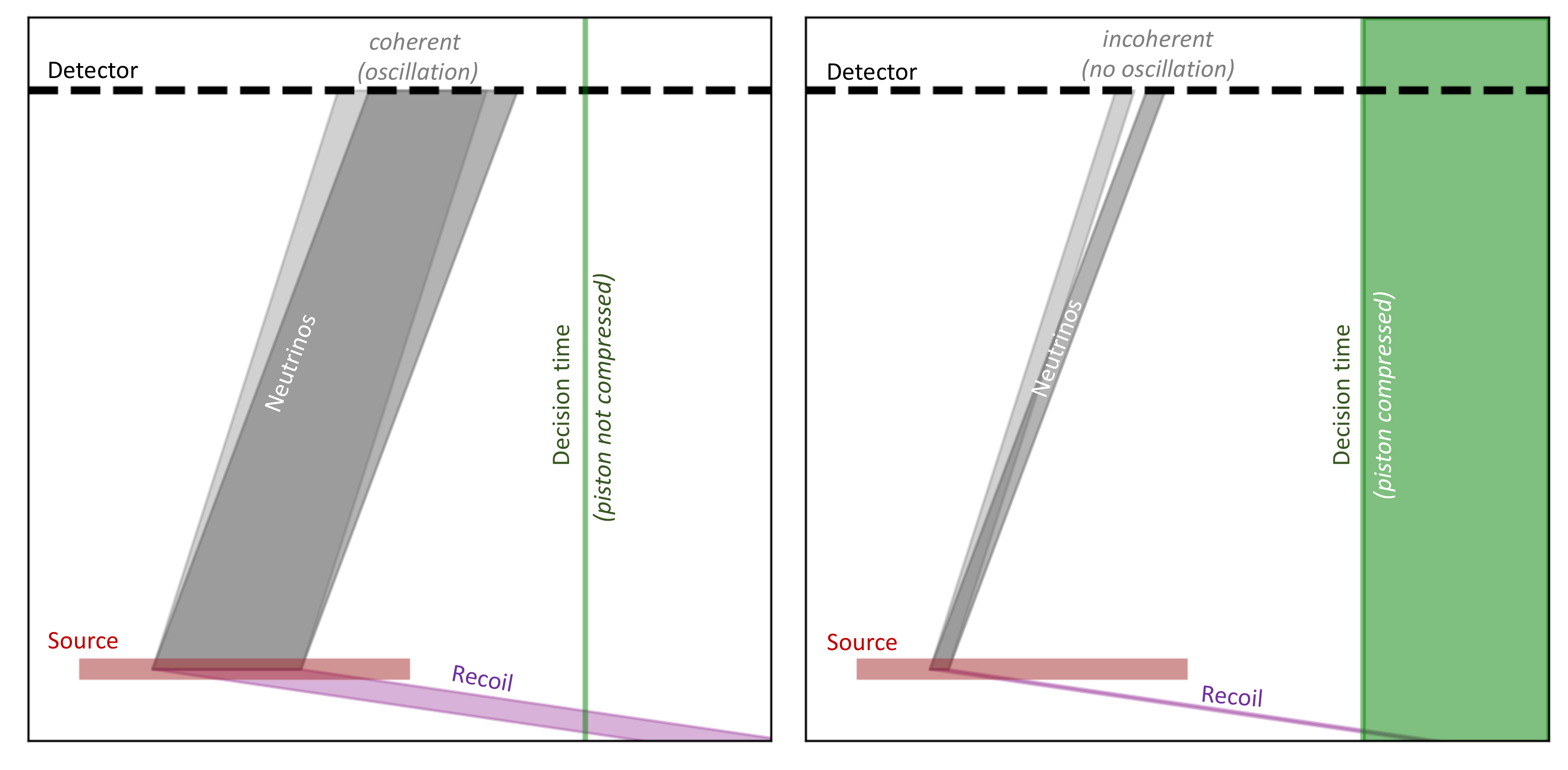}
\par\end{centering}
\caption{Thought experiment where the formalism of Ref.~\cite{akhsmirn} implies
backward propagation of information in time. The left panel shows
the effect of not compressing the piston; the right of compressing
it. According to the formalism of Ref.~\cite{akhsmirn}, this mechanism
can be used to send information backwards in time via the neutrino
oscillation probability. All omitted labels are equivalent to those in Fig.~\ref{fig:Light-cone-of}. \label{fig:Thought-experiment-where}}
\end{figure}

To make this argument less abstract, let us contrive of an unrealistic
but physically consistent thought experiment where superluminal signaling
would be manifest, were the approach of Ref.~\cite{akhsmirn} correct.
We consider a gaseous source that generates neutrinos via electron
capture, and contains a sufficient number atoms that a high flux of
neutrinos is continuously being emitted. The choice of electron capture
for this example means we can ignore the effects of an emitted electron
and focus only on the daughter nucleus. We collect enough neutrinos
with an idealized detector to monitor the oscillation probability
at $L$ at any time. We can control the source density $\rho$ by
compressing the gas with a piston. Our procedure will be as follows:
\begin{enumerate}
\item Start the experiment with the gas at a sufficiently low density that
interactions of the final state nuclei do not cause any observable
decoherence effects (Fig. \ref{fig:Thought-experiment-where}, left).
\item At some moment $t_{0}$, we make a decision: either compress the piston
leading to an increase in the gas density, or leave the gas in its
present low-density state. If we compress the piston, the density
$\rho(t>t_{0})$ becomes high enough to induce significant wave-packet
separation effects on the oscillation probability for neutrinos at
distance $L$ (Fig. \ref{fig:Thought-experiment-where}, right). 
\item By choosing to compress or not compress the cylinder we may send a
bit of information via the neutrino oscillation probability. We ask:
at what time does this bit of information arrive?
\end{enumerate}
The formalism of Ref.~\cite{akhsmirn} provides an answer to this
question. Neutrinos detected at time $t'$ were emitted at $t'-\frac{L}{c}+{\cal O}(\frac{m}{E})$.
Since we choose to begin our experiment with gas at a sufficiently
low density that the mean-free time $t_{N'}\gg\frac{L}{c}$, the recoils
are still travelling even after the neutrinos have been detected. 

Upon compressing the piston at time $t_{0}$, we transition to the
incoherent mode. If the interactions of the entangled recoil truly
impact the oscillation probability, the relevant neutrinos that are
decohered were those emitted at $t_{0}-t_{N'}$, and then detected
at $t_{D}=t_{0}-t_{N'}+\frac{L}{c}+{\cal O}(\frac{m}{E})$. Since
$t_{N'}\gg\frac{L}{c}$, the detection time $t_{D}<t_{0}$. If we
measure the oscillation probability of these neutrinos and find it
influenced by the interactions of the final state nuclear recoil,
we have succeeded in sending a signal backwards in time.

\section{Choice of basis for tracing out an entangled partner, and analogies
to the EPR experiment \label{sec:Analogies-to-and}}

The system as described has striking similarities to the original
EPR experiment \cite{einstein1935can}. Here we explore some similarities
and differences, and this ultimately proves enlightening as to the
origin of the apparent causality violation of Ref.~\cite{akhsmirn}. Some of our treatment here resembles the discussion in Appendix A of  Ref.~\cite{jones2015dynamical}.

In the original EPR experiment, two qubits $A$ and $B$ are prepared
in an entangled state and moved to a large spatial separation:
\begin{equation}
|\psi\rangle=\frac{|\uparrow_{Z}\rangle_{A}\varotimes|\downarrow_{Z}\rangle_{B}+|\downarrow_{Z}\rangle_{A}\varotimes|\uparrow_{Z}\rangle_{B}}{\sqrt{2}},\label{eq:EPRState}
\end{equation}\
Measurement of system $A$ influences system $B$ via wave function
collapse. For example, if $A$ measures spin in the $\hat{z}$ direction
and obtains $\downarrow_{Z}$, the wave function of $B$ becomes:
\begin{equation}
|\psi\rangle_{B}\rightarrow|\uparrow_{Z}\rangle_{B}.
\end{equation}
This final state for $B$ is only possible if $A$ has measured along
$\hat{z}$. Conversely, if $A$ measures the spin along $\hat{x}$
direction and obtains $\downarrow_{X}$, the $B$ wave-function would
become:
\begin{equation}
|\psi\rangle_{B}\rightarrow|\uparrow_{X}\rangle_{B}=\frac{|\uparrow_{Z}\rangle_{B}+|\downarrow_{Z}\rangle_{B}}{\sqrt{2}}.
\end{equation}
It seems therefore that $A$ can encode information in $B$'s spin
at a large physical distance by choosing what quantity to measure.
This ``spooky action at a distance'' invited arguments against the
plausibility of quantum mechanics, which were the motivation for Einstein,
Podolsky and Rosen's seminal analysis. The original EPR discussion
resolved how this apparent information transfer can be reconciled
with requirements of causality. The resolution to this foundational
question is that $B$ cannot access the newly encoded information
without an additional, classical signal sent from $A$ to $B$, which
must by definition travel slower than light. This is because no matter
what $A$ has done or not done to her qubit, given an initial state
as in Eq.~\ref{eq:EPRState}, $B$ will find a 50\% probability of
spin up as long as he measures his qubit alone while ignoring the
outcome of $A$'s measurements. Only in cases where correlation measurements
are performed that compare outcomes of measurements at $A$ to those
at $B$ can the non-trivial consequences of quantum entanglement be
observed.

This is easy to prove. We posit that $A$ makes some measurement defined
by Hermitian operator $M$, with possible outcomes $|\lambda_{i}\rangle$
for i=1,2, where $M|\lambda_{i}\rangle=\lambda_{i}|\lambda_{i}\rangle$.
The probability for $B$ to find an answer $|\uparrow\rangle$ is
given by a probability-weighted sum over outcomes of $M$ on $A,$
\begin{equation}
P(\uparrow_{B})=\sum_{i}\left(\langle\lambda_{i}|\varotimes\langle\uparrow_{B}|\right)|\psi\rangle\langle\psi|\left(|\lambda_{i}\rangle\varotimes|\uparrow_{B}\rangle\right).\label{eq:ProbOfUp}
\end{equation}
The mathematics is least tedious in the density matrix picture, though
an identical proof can be completed with wave functions. We introduce
the bipartite density matrix for the $AB$ system $\rho=|\psi\rangle\langle\psi|$,
and the reduced density matrix for system $B$ is obtained by ``tracing
out'' the entangled $A$ subsystem:
\begin{equation}
\rho_{B}=\sum_{i}\langle\lambda_{i}|\psi\rangle\langle\psi|\lambda_{i}\rangle\equiv\mathrm{Tr}_{A}\left[\rho\right].
\end{equation}
The probability Eq.~\ref{eq:ProbOfUp} for B to obtain $\uparrow$
is then conveniently expressed as
\begin{equation}
P(\uparrow_{B})=\langle\uparrow_{B}|\rho_{B}|\uparrow_{B}\rangle.
\end{equation}
The probability for any other measurement we can imagine $B$ making
on his qubit alone can be found analogously, using $\rho_{B}$. The
construction accounts implicitly for the sum over outcomes of measurement
$M$ on $A$.

The important principle that ultimately leads to resolution of the
EPR paradox is that the reduced density matrix $\rho_{B}$ is independent
of which $M$ has been made on $A$. For a different $M$, we would
have different measurement outcomes, related to the original ones
by $|\lambda_{i}'\rangle=U_{ij}|\lambda_{j}\rangle$ for some unitary $U$.
The new reduced density matrix is $\rho'_{B}$:
\begin{equation}
\rho'_{B}=\langle\lambda_{i}|U^{\dagger}|\psi\rangle\langle\psi|U|\lambda_{i}\rangle.
\end{equation}
Inserting two complete sets of states into this expression we find:
\begin{eqnarray}
\rho'_{B}&=&\langle\lambda_{i}|U^{\dagger}|\lambda_{a}\rangle\langle\lambda_{a}|\psi\rangle\langle\psi|\lambda_{b}\rangle\langle\lambda_{b}|U|\lambda_{i}\rangle\\
&=&U_{bi}U_{ia}^{\dagger}\langle\lambda_{a}|\psi\rangle\langle\psi|\lambda_{b}\rangle\\
&=&\delta_{ba}\langle\lambda_{a}|\psi\rangle\langle\psi|\lambda_{b}\rangle\\
&=&\rho_{B},
\end{eqnarray}
and so, no matter what $A$ does or does not do to her qubit, measurements
made by $B$ on his entangled qubit without knowledge of the outcome
of $A$'s manipulations will have the same probabilities. This prohibits
superluminal signaling, and reconciles the instantaneous nature of
wave function collapse with the requirements of causality. This conclusion
derives from properly summing $B$'s predictions over the probabilistic
outcomes of $A$'s measurement (whatever they may be), as realized
in the partial trace operation.

How is the neutrino-recoil system similar or different to the EPR
system? We can replace $A$ with the neutrino, $B$ with the recoil,
and move from 2D Hilbert spaces to continuous ones. Any measurement
of the neutrino alone (say, its flavor $\beta$ at baseline $L$)
can be made by probabilistically summing over possible outcomes for
the recoil, which we express in a basis $|\lambda\rangle$ for some
continuous $\lambda$:

\begin{equation}
P(\beta;L)=\int d\lambda\left(\langle\lambda|\varotimes\langle\beta,L|\right)|\psi\rangle\langle\psi|\left(|\lambda\rangle\varotimes|\beta,L\rangle\right)\\
=\langle\beta,L|\rho_{\nu}|\beta,L\rangle,
\end{equation}
where the neutrino reduced density matrix is
\begin{equation}
\rho_{\nu}=\mathrm{Tr}_{\lambda}[\rho]=\int d\lambda\langle\lambda|\rho|\lambda\rangle.
\end{equation}

By the same argument as above, the neutrino reduced density matrix
$\rho_{\nu}$ is independent of what we choose to measure about the
recoil, since $\rho'_{\nu}$ for any choice of recoil basis states
$U|\lambda\rangle$ is independent of $U$. It is also independent
of the subsequent time evolution of the recoil, because the time evolution
operator for the recoil is itself just another unitary operation:
\begin{equation}
|\lambda(t)\rangle=U_{R}(t-t_0)|\lambda\rangle.
\end{equation}
If we change the nature of the time evolution of the recoil (for example
by changing the medium into which it is emitted), we then switch out
the $U_{R}$ operator for a different one. According to the above
proof, $\rho_{\nu}$ must be unaffected and the neutrino oscillation
probability must remain unchanged. The conclusion is that neither
the time evolution nor the measurement of the recoil can influence
the overall oscillation probability of the neutrino as long as we
do not make correlation measurements - just as in the EPR experiment.

How then do we reconcile this with the proofs of Ref.~\cite{akhsmirn}
that suggest that the total oscillation probability is indeed influenced
by the subsequent interactions of the recoil? To be maximally explicit,
the claim there is that beyond some distance $L_{coh},$ oscillations
of the neutrino alone would be observable were the recoil emitted
into empty space, but unobservable if it were emitted into sufficiently
dense matter, thus introducing an explicit dependence of $\rho_{\nu}$
on the operator $U_{R}(t-t_0)$.

This problem seems to be introduced by choosing to project the recoil
onto a Gaussian wave-packet. Notably such functions do not represent
an orthonormal basis of final states for the recoil. As such, the
process of determining which Gaussian wave packet the recoil lands
in is not a measurement in the sense described by the postulates of
quantum mechanics. There, measurements are supposed to be described
by Hermitian operators, whose eigenstates are necessarily orthogonal.
The problem here is that taking a sum over sets of non-orthogonal
final states is not a valid way to probabilistically sum the possible
fates of the recoil, and this introduces unphysical effects into the
oscillation probability prediction.

We note that it is possible to introduce this same malady in the original
EPR scenario by using non-orthogonal measurement states. Imagine that
we do not force $A$ to make a measurement in the strict quantum mechanical
sense, but instead allow $A$ to ask, which of the two (non-orthogonal)
states below is her qubit found in?
\begin{eqnarray}
|\mu_{1}\rangle&=&|\uparrow\rangle,\\
|\mu_{2}\rangle&=&\alpha|\uparrow\rangle+\sqrt{1-\alpha^{2}}|\downarrow\rangle,\quad\quad0\leq\alpha\leq1.
\end{eqnarray}

If we now we sum $B$'s probabilities for spin up over $A$'s probabilities
for the outcomes $\mu_{1}$ and $\mu_{2}$, we find
\begin{eqnarray}
P(\uparrow_{B})&=&\sum_{i}\left(\langle\mu_{i}|\varotimes\langle\uparrow_{B}|\right)|\psi\rangle\langle\psi|\left(|\mu_{i}\rangle\varotimes|\uparrow_{B}\rangle\right)\label{eq:SumPbTilde}\\
&=&\langle\uparrow_{B}|\tilde{\rho}_{B}|\uparrow_{B}\rangle,\label{eq:ProbB}
\end{eqnarray}
The object $\tilde{\rho}_{B}$ is not a reduced density matrix,
but some analogous object constructed with non-orthogonal basis states,
\begin{equation}
\tilde{\rho}_{B}=\langle\mu_{i}|\psi\rangle\langle\psi|\mu_{i}\rangle.
\end{equation}
The matrix $X$ that relates the new basis states to the original
ones is now non-unitary unless $\alpha=0$:
\begin{equation}
|\mu_{i}\rangle=X(\alpha)_{ij}|\lambda_{j}\rangle,\quad\quad X(\alpha)X^{\dagger}(\alpha)=\left(\begin{array}{cc}
1+\alpha^{2} & \alpha\sqrt{1-\alpha^{2}}\\
\alpha\sqrt{1-\alpha^{2}} & 1-\alpha^{2}
\end{array}\right).
\end{equation}
As a result we find
\begin{eqnarray}
\tilde{\rho}_{B}&=&X_{bi}(\alpha)X_{ia}^{\dagger}(\alpha)\langle\lambda_{a}|\psi\rangle\langle\psi|\lambda_{b}\rangle
\\
&\neq&\rho_{B},
\end{eqnarray}
which now depends on $\alpha$. This dependence on $\alpha$ is also
manifest in the probabilities $P(\uparrow_{B}$) of Eq.~\ref{eq:ProbB}.
As such, superluminal communication between $A$ and $B$ now appears
to be allowed by suitable choice of $\alpha$ by $A$.

This apparent causality violation emerges from mistaken assumption
that Eq.~\ref{eq:SumPbTilde} may be considered as a sum over probabilistic outcomes at $A$. There is no quantum mechanical measurement that
$A$ could make that would have possible outcomes $|\mu_{1}\rangle$
and $|\mu_{2}\rangle$, since they are not orthogonal states. As a
consequence, summing over their probabilities is an improper construction
to integrate over possible final states of the distant $A$ system.
This means the probabilities calculated for the measurements at $B$
are invalid. The origin of the apparent causality violation in the
neutrino-recoil system appears to be the same as in this toy example. 

\section{Localization through interaction with the environment\label{sec:Localization-through-interaction}}

The above examples illustrate that the neutrino oscillation probability
cannot possibly depend on the interactions of the recoiling nucleus
after it leaves the decay, or on the interactions of any other entangled
daughter particles with surrounding material. 

On the other hand, that the entangled final-state particles do both
exist and carry away information about the neutrino away from the
decay does impact neutrino coherence properties. Such information
escapes into the universe recording information about the neutrino
whether they are observed or not, and suppresses oscillation coherence
in certain circumstances. The key point is that, while it is always
important that that recoil particle(s) were emitted in the decay,
what happen to them after production cannot influence the neutrino-only
oscillation phenomenology, without violating causality. Tracing over
the recoil final states in any orthogonal basis will generate the
same oscillation probabilities for the neutrino, and this appears
to be the the only well defined way to probabilistically sum over
final states of the recoil.

While environmental interactions of the daughter particles cannot
influence coherence, those of the parent can and do. Entanglements
generated between the parent and others through collisions serve (colloquially)
to ``measure'' the position of the emitter prior to its decay with
a precision dictated by the momentum transfer in the interaction \cite{tegmark1993apparent}.
Those interactions are in the past light cone of the neutrino for
every observer, so no causality problems are implied by their influence.
The Weisskopf-Van-Vleck formalism of Ref.~\cite{parsons1968collision}
as cited in Ref.~\cite{akhsmirn} in fact describes the effects of
these initial-state interactions on determining line shapes, and not
those of the final-state recoil after the decay. 

The question of which interactions do serve to measure the parent
position appears to be not as clear-cut as described in Ref.~\cite{akhsmirn},
and there remains ambiguity in the relevant distance scale. While
atomic collisions are advanced there as the relevant localizations,
we may also consider neutrino production as proceeding via the process
$n\rightarrow p+e+\nu_{e}$ for neutrons bound inside a nucleus. Thus
another reasonable localization scale appears to be that induced by
the interaction among nucleons inside the nucleus itself. After the
decay the remaining N-1 entangled nucleons are left behind, and their
position encodes the location of the original emitter, localizing
it to something of order the size of the nucleus itself. This would
seem to imply a far smaller wave packet width and more dramatic decoherence
effects than the modest localization scale from atomic collisions
of either the parent or daughter particles. A detailed treatment of
the implied oscillation phenomenology is outside the scope of the
present comment, but is the subject of a forthcoming paper.\bibliographystyle{unsrt}

\section*{Acknowledgements}

We thank Alexei Smirnov and Evgeny Akhmedov for enlightening discussions that helped to sharpen the arguments given in this comment. We thank Joshua Spitz, Eric Marzec, Carlos Arg\"uelles, Saori Pastore, Raquel Castillo Fernandez, Krishan Mistry and Grant Parker for their comments on the manuscipt. BJPJ is supported by Department of Energy under Award numbers {DE-SC0019054} and {DE-SC0019223} and the National Science Foundation via the IceCube South Pole Neutrino Observatory. 

\bibliography{main}

\end{document}